\documentclass{iau_FM}
\usepackage{graphicx}

\title[Star Formation Close to Sgr A*] 
{Star Formation Close to Sgr A* and \\
Beyond the Nuclear Cluster}

\author[Yusef-Zadeh \& Wardle]   
{Farhad Yusef-Zadeh$^1$
 \and Mark  Wardle$^2$}

\affiliation{
$^1$CIERA, Department of Physics and Astronomy, Northwestern University, \\
2145 Sheridan Rd, Evanston, IL 60208, US \\email: {\tt zadeh@northwestern.edu}\\
$^2$Department of Physics and Astronomy and
Research Center for Astronomy, \\Astrophysics \&  Astrophotonics, Macquarie University,
Sydney NSW 2109, Australia \\ email: {\tt mark.wardle@mq.edu.au} \\[\affilskip]
}

\pubyear{2016}
\jname{International Astronomical Union Proceedings Series} 
\editors{Andreja Gomboc and  Carole Mundell}
\begin{document}

\maketitle

\begin{abstract}
Two modes of star formation are involved to explain the origin of young
stars near Sgr A*. 
One is a disk-based mode,  which explains the disk of stars orbiting Sgr A*. 
The other is the standard cloud-based mode 
observed in the Galactic
disk. 
We  discuss each of these modes of star formation and apply these ideas to the  inner few parsecs 
of Sgr A*.  In particular, we  focus on the latter mode  in more detail. 
We also discuss how the tidal force exerted by the nuclear cluster
makes  the Roche density approaching zero  and  
contributes   to the collapse of molecular clouds located tens of parsecs
away from Sgr A*. 
\end{abstract}
\firstsection 
\section{Introduction}

The environment of Sgr A*, the 4 million Solar mass  black hole at the center of the 
Galaxy,  provides   a window to close-up  study of the center of galaxies and is a   
potential  Rosetta Stone for understanding  star formation  under extreme
physical  conditions.  A critical question is whether tidal shear in the vicinity of 
supermassive black holes (SMBHs) is able to completely suppress or induce star
formation. 

Star formation is modified 
by tidal  effects due to Sgr A*  
as well as high external pressure due to variety of processes  close to Sgr A*. 
There is  empirical evidence for three 
different regimes of  star formation in this region of the Galaxy, as described below. 

First, close  to Sgr A*, there is a concentration 
of over 100 OB stars  within 0.04 and 0.4 pc of Sgr A*
providing  strong evidence for  formation of stars 
within  0.5 pc of  Sgr~A* in the last few million years (Paumard et al. 2006; Lu et al. 2009). 
 To explain star formation in a disk,  a plausible suggestion is 
that gravitational collapse took place in a gaseous disk captured by the black hole. 
Star formation in a captured disk around Sgr A* explains the compactness of
the observed stellar disk. 
This implies  that the raw material for the disk
was captured from a cloud as it temporarily engulfed Sgr A* several million
years ago while passing through the central parsec of the Galaxy. 
Due to the cancellation of angular momentum of the captured cloud
material that passes on opposite sides of the black hole, it  naturally produces
a compact, gravitationally unstable disk (Wardle and Yusef-Zadeh 2008, 2014). 
One of the key feature of this mechanism  is that clouds of gas can be 
brought in very close to supermassive black holes. This is mainly because 
of the loss of angular momentum of the initial cloud,  
thus  the captured cloud generates  not only 
a high rate of star formation  but also provides   the fuel necessary to feed  SMBHs.


\begin{figure}[b] 
 \includegraphics[width=3.8in]{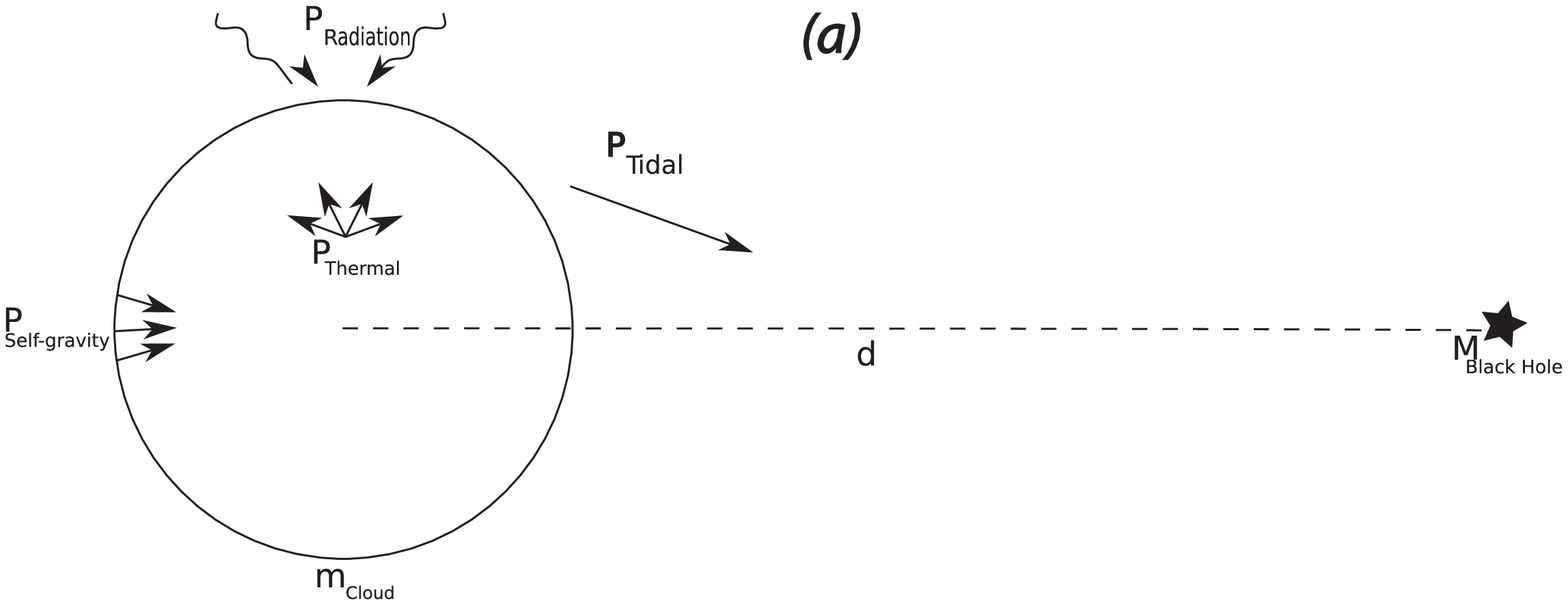}
\begin{center}
 \includegraphics[width=4.9in]{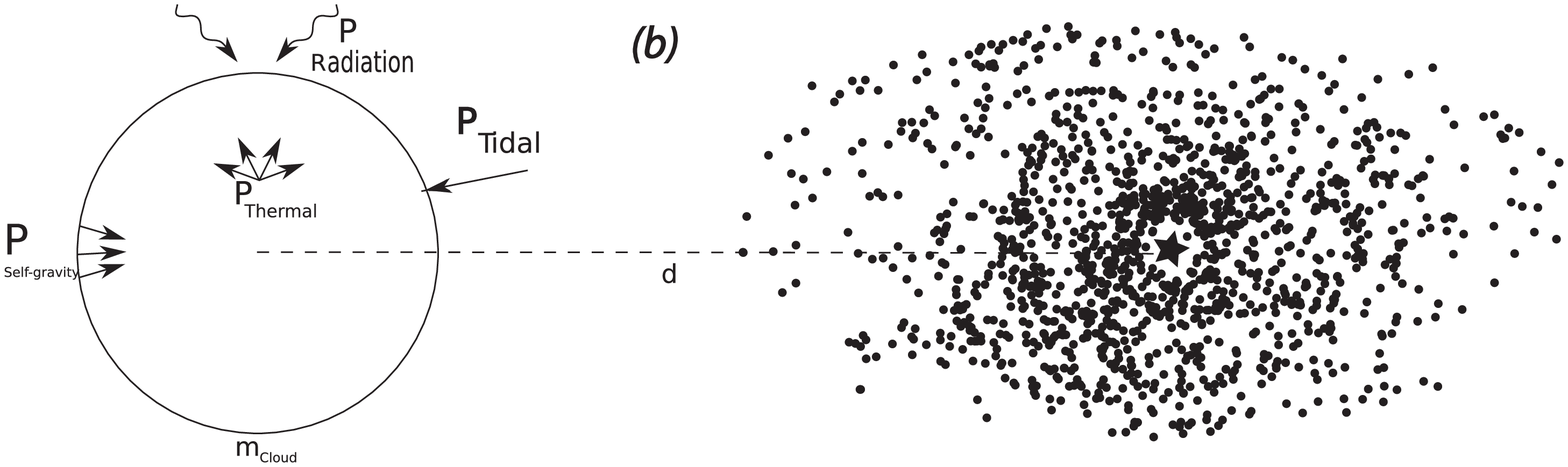} 
 \caption{
{\it (a) Top} 
A Schematic diagram showing the four forces that a cloud experiences as it approaches  Sgr A*
with its point mass gravitational potential.  
{\it (b) Bottom}  Same as (a) except that an extended concentration of mass (the nuclear cluster) 
serves as the gravitational potential and the tidal force on the cloud becomes  compressive.
}
\end{center}
\end{figure}


Second, massive   stars 
such as IRS 8 and about ten 
bow-shock early type stars are distributed within the ionized streamers
0.5-2 pc  from Sgr A*  (Geballe et al. 2006; Sanchez-Bermudez et al. 2014).    
On a scale of tens of parsecs, 
there  are  also  clusters of massive 
young stars (e.g., the Arches  and Quintuplet star clusters) 
with similar ages to those  of the central cluster near Sgr A* (Stolte et al. 2015).
The question is how star formation proceeds in these three different regions.


In addition to the  population  of young massive stars near Sgr A*, 
several lines of evidence point to on-going star formation 
within two  parsecs of Sgr A*. 
First,  the discovery of  water masers with multiple and single velocity components 
(Yusef-Zadeh et al. 2015a). Second, 
SED modeling of 64 infrared excess sources in the inner pc of Sgr A* 
are consistent with their being 
YSOs (Yusef-Zadeh et al. 2015a). 
Third, 44 
partially resolved compact sources with size scales ranging between 400 and 1600 AUs have been identified
(Yusef-Zadeh et al. 2015b). 
The bow-shock appearance of these sources facing  the direction of Sgr A* suggests
a  population of 
photoevaporative protoplanetary disks (proplyds) 
associated with newly formed low 
mass stars. 
The disks are externally illuminated by strong Lyman continuum radiation from the 100 OB and WR massive 
stars distributed within 0.5 pc of Sgr A*. We have recently detected millimeter emission from five 
proplyd candidates supporting the idea that these sources have 
cool disks (Yusef-Zadeh et al. 2016; submitted).  
Fourth, several SiO (5-4) sources   within a pc of Sgr A*
are suggestive of YSOs with protostellar outflows.
Their SiO (5-4) luminosities and their velocity widths  are similar to 
protostellar outflows in star forming clouds in the disk 
of the Galaxy (Yusef-Zadeh et al. 2013).  
Finally,  a  recent ALMA study of this region (Tsuboi et al. 2016) 
suggests  molecular clumps of 10-100 Solar mass,
distributed in the streamers.
Altogether the evidence for on-going star formation implies that 
stellar disk of OB stars is  formed close to Sgr A*  rather than being migrated from large distances. 
Here, we focus 
on the mode of star formation beyond the inner 0.5 pc of Sgr A* and 
summarize  the results of a recent work, 
details of which   can be found in  Wardle and Yusef-Zadeh (2016, in preparation).



\begin{figure}[b] 
\begin{center}
 \includegraphics[width=3.5in]{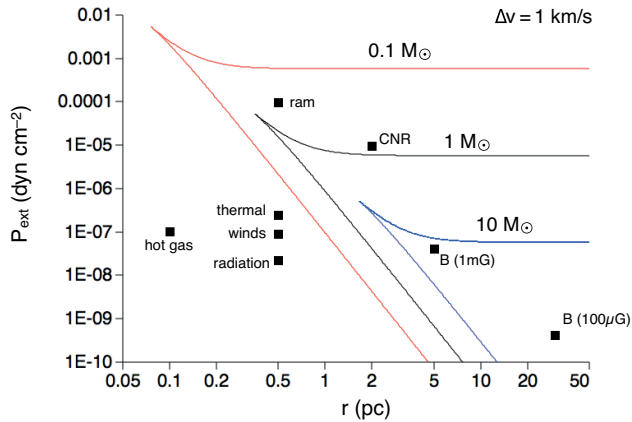}
\end{center}
 \caption{ 
The external pressure experienced by a Galactic center cloud as 
a function of distance. 
The plots in the top, middle  and bottom  correspond to 
 0.1, 1 and 10 Solar mass  cloud masses, respectively.   
}
\end{figure}

\section{Cloud-based Mode of Star Formation}
Although  cloud capture model can explain  the origin of the stellar disk orbiting Sgr~A*, 
a different process is needed to explain the formation of isolated stars or of clusters 
 beyond 0.5 pc of Sgr A*. The gravitational potential within the inner $\sim$5 pc is 
dominated by the mass of  Sgr A*. 
The Roche density is  10$^8$ cm$^{-3}$ at 1 pc from Sgr A* and the gas density inferred 
from a number of measurements  is lower by 
at least  1.5 orders  of magnitude.  
However, the  gravitational stability of a cloud 
also depends on the  internal and external pressures 
(Chen et al. 2016; Wardle and Yusef-Zadeh 2016). 
Figure~1a demonstrates the  disruptive forces due to 
the tidal force from Sgr A*  
and the expansive  internal pressure. 
These are opposed by confining forces, 
self-gravity and  compression by external pressure such as  
radiation,  cosmic rays  and   shocks.  
Thus,  the stability and the evolution 
of a cloud clearly depends on the balance between the internal and external forces.      

Figure 2 shows the range of external pressures for which clouds with 1 km s$^{-1}$ line width are stable as a function 
of distance from Sgr A*.  There are different types of pressure forces acting on a cloud at different distance from 
Sgr A*. The calculated value of pressure on each curve 
should be compared to the actual pressure at each radius, which is contributed by a variety of 
different mechanisms such as the ram pressure associated with motion through the external medium or the pressure 
exerted by a jet or outflow from Sgr A* that impinges upon the cloud. 
The circumnuclear molecular ring with its high gas density, 
the hot X-ray gas, stellar winds from massive stars, 
the UV radiation from massive stars,  the ionized thermal  gas in the mini-spiral (the streamers) and  
the magnetic field  contribute within a few parsecs of Sgr A*. 

By way of illustration consider the fate of 
a one  Solar mass  cloud. Figure  2 
shows  that this cloud could collapse if it is at least 
 0.5 pc from Sgr A*.  
If the cloud is closer than this, 
 it will be torn apart by tidal effects. 
In the region between the diagonal and horizontal boundaries, 
 the cloud remains stable and neither collapses nor 
is disrupted.  
One of the important implications is that lower mass clouds can 
approach Sgr A* more closely than high mass clouds. 
 Given sufficient external pressure, 
lower mass stars are preferentially formed 
 closer to the black hole.  This implies that 
a  population of 
proplyds
associated with newly formed low 
mass stars should be in the vicinity of Sgr A* (see Yusef-Zadeh et al. 2015b).  



\section{Star Formation Beyond the Nuclear Cluster}
SMBHs in external galaxies with a mass $\sim100-1000$ times Sgr A*, dominate 
the nuclear stellar potential within 100 pc to 1 kpc. 
Star formation in the innermost regions of galaxies is generally
presumed to be suppressed by strong tidal forces. 
In the case of Sgr~A*, the potential on 10-200 pc scales is dominated 
by the nuclear cluster. 
It is generally presumed (e.g. Morris \& Serabyn 1996) that tidal forces can only be overcome if the gas density exceeds
the Roche density based on the interior mass. However, this 
condition  may never arise since  
the Roche density drops to zero wherever  the gravitational potential 
is not Keplerian, as 
described below.


It turns out that tidal forces respond differently when a cloud is experiencing a potential 
dominated by a point mass, Sgr A*,   vs an extended mass concentration 
(e.g., the nuclear cluster (Jog 2013; Wardle \& Yusef-Zadeh 2016).  The schematic diagram in Figure 1b presents 
the  distribution of the nuclear cluster  centered on Sgr A*. 
A cloud within a couple of parsecs experiences mainly the 
Keplerian potential of Sgr A* and at large distances beyond a few parsecs, the potential of 
the nuclear cluster becomes significant. 
The rotation curve is  Keplerian in the inner parsec, where the enclosed mass 
is dominated by Sgr A*, flattens between 1 and 10\,pc as the stellar cluster
starts to dominate the enclosed mass, and slowly rises beyond that.  
The consequence of such a change in the mass distribution profile is 
that   clouds latteral compression by 
the tidal force of the nuclear cluster dominates radial stretching. 
Another consequence  of this external  force on a cloud 
beyond the nuclear cluster is that the   Roche density approaches zero. 
The reason for such a drop in Roche density  is that 
the differential force across a cloud 
is not significant in the radial direction as long as the rotation curve is flat or rising. 
Thus,  the acceleration across 
the radial direction becomes insignificant. 
This situation is different for a cloud embedded in a potential of  a point mass 
where the rotation curve  drops rapidly as a function 
of distance from Sgr A*. 
This result is different than  what has generally been assumed 
that the gas density must  be 
greater than the Roche density. It is conceivable that prominent 
young star clusters such as the Arches and Quituplet clusters benefited from the 
contribution of the nuclear cluster in compressing the gas cloud, thus the cloud becomes gravitationally 
unstable.  If the potential is dominated by the Dark matter in the early universe, 
then star formation may be induced in a cloud as long as clouds of gas do not follow a point mass potential.


\end{document}